\documentclass[amsmath,amssymb,prl,showpacs,twocolumn]{revtex4}

\usepackage{graphicx}

\begin{document}

\title{Charge instabilities at the metamagnetic transition}
\author{Carsten Honerkamp}
\affiliation{Max Planck Institute for Solid State Research, D-70569 Stuttgart, 
Germany}
\date{February 16, 2005}

\begin{abstract}
We investigate instabilities in the charge channel in the vicinity 
of (meta-)magnetic transitions of itinerant electron 
systems. Based on a weak coupling analysis we argue that
in a one-band $t$-$t'$ Hubbard model near the van Hove filling and dominant 
ferromagnetic fluctuations it is difficult to account for a microscopic 
mechanism for a $d$-wave Pomeranchuk deformation of the Fermi surface. 
A similar  deformation has been considered for 
the metamagnetic transition in Sr$_3$Ru$_2$O$_7$.
As an alternative we discuss the possibility of 
charge inhomogeneity on the nano scale. 
This extends the analogy of the metamagnetic transition to a liquid-gas transition.
\end{abstract}


\maketitle

Over the last decade, the perovskite strontium-ruthenates have become a model system for the study of
correlated electrons. 
While the infinite-layer compound  SrRuO$_3$ is an itinerant ferromagnet\cite{cao} with Curie temperature of $T_C=165$K, the single-layer Sr$_2$RuO$_4$ is a triplet superconductor\cite{sc214} below $T_c=1.5$K. Above $T_c$ it is a 
well-studied strongly anisotropic Fermi liquid\cite{bergemann}. 
By substituting Sr with La and thereby enhancing the Fermi surface (FS) volumes, it can be brought into the vicinity of 
ferromagnetism (FM)\cite{kikugawa}. 
The double-layered Sr$_3$Ru$_2$O$_7$ interpolates between these two limiting cases in an exciting way. 
While first experiments found FM, subsequent research\cite{ikeda} made clear that the magnetization $M$ is zero in 
the absence of a magnetic field $B$.
However at a direction-dependent critical field  strength $B_m$ between 4 and 8$T$, $M(B)$ shows a steep increase below 
$T\sim 2$K, referred to as a metamagnetic transition\cite{perry01,grigera01}. 
Around $B_m$, there are indications for non-Fermi liquid behavior, e.g. in the specific heat and the $T$-dependence 
of the inplane-resistivity\cite{perry01,grigera01}. 
Theoretically, the metamagnetic transition with a single jump in the magnetization can be understood  in a 
Stoner-type model with a FS near a van Hove (VH) singularity in the density of states\cite{binz,memag}. 
Indeed a number of experiments\cite{bergemann} for the single-layer system  Sr$_2$RuO$_4$ and band structure calculations\cite{singh} 
for both Sr$_2$RuO$_4$ and Sr$_3$Ru$_2$O$_7$ confirm the presence of a nearby van Hove singularity for the FS derived from the Ru $d_{xy}$-orbitals.

A new twist occurred when improved sample quality revealed at least two jumps\cite{ohmichi,perry04} in the magnetization when $B$ is increased. 
Most interestingly, 
in the critical field range between the two jumps of Ref.~\cite{perry04}, the 
residual resistivity $\rho_0$ is about twice as high as below and above the 
jumps. 
As a function of the temperature, $\rho(T)$ saturates already at $T= 1$K in the critical field range, 
and the $T$-dependence above is of non-Fermi-liquid type. 
The high $\rho_0$ can be understood if one 
assumes the formation of domain-like structures in the critical range\cite{grigera04}. Recently, two works\cite{grigera04,kee} suggested the occurrence of a $d_{x^2-y^2}$-wave-like Fermi surface deformation ($d$FSD). In a one-band picture on the two-dimensional (2D) square lattice, the $d$FSD splits the two inequivalent VH 
points and thereby gains energy if the forward scattering in the charge channel has an attractive $d$-wave component. 
This possibility has been analyzed for 
Hubbard-like models by a number of authors\cite{yamasekohno,halboth,neumayr,yamase,schofield}. 
An experimental realization of the effect has not been found yet. 
For the double-layer ruthenate this picture is however quite 
appealing\cite{grigera04}. The occurrence of two 
magnetization jumps can 
be understood in the $d$FSD framework as first-order entry into and exit out 
of the phase with reduced FS symmetry\cite{kee}. Moreover the 
sensitivity of the experimental picture with respect to sample quality seems to 
fit into the scenario of an unconventional order parameter. 
We note that while the chemical-potential-tuned 
quantum phase transition into the $d$FSD state is typically first order, 
the thermal transition at $T_c$ is second order over a wide parameter range 
in the simple forward scattering model\cite{khavkine,yamaseS}. 
 
So far the microscopic origin of an attractive $d$-wave component in the forward scattering has not been addressed 
in the context of the metamagnetic transition. In this work we analyze this question based on the one-band $t$-$t'$ Hubbard model on the 2D square lattice and specify the parameter region in which the $d$FSD is favorable. 
Furthermore we reinvestigate the Stoner model for the metamagnetic transition\cite{binz}. We point out that 
in this model, the magnetization jump is located in an unstable density region. 
This extends the analogy to a liquid-gas transition already emphasized by Green et al.\cite{green}. 
Around the transition charge inhomogeneity might occur and give rise to a higher residual resistivity. 
We also describe a possibility how more than one jump in the 
magnetization can occur in this model. 

First we consider the effective forward scattering in the charge channel 
for the 2D $t$-$t'$ Hubbard model\cite{ttp} on the square lattice in zero magnetic field. A perturbative but unbiased way to arrive at effective low-energy or low-$T$ interactions is the functional 
renormalization group (fRG). In fact, the $d$FSD or Pomeranchuk tendencies in the 2D Hubbard model were revealed using fRG\cite{halboth}. 
Here, we use the temperature-flow fRG\cite{tflow}. 
It has the additional advantage of not being blind with respect to long-wave length particle-hole fluctuations. 
Using this method, a low-$T$ $p$-wave superconducting instability was found for a FS similar to that of 
the $\gamma$-band in Sr$_2$RuO$_4$\cite{tflow}. A second order transition into a state with broken symmetry is indicated by a 
divergence of the corresponding susceptibility at $T_c$ when $T$ is lowered.

\begin{figure} 
%\begin{center} 
\includegraphics[width=.47\textwidth]{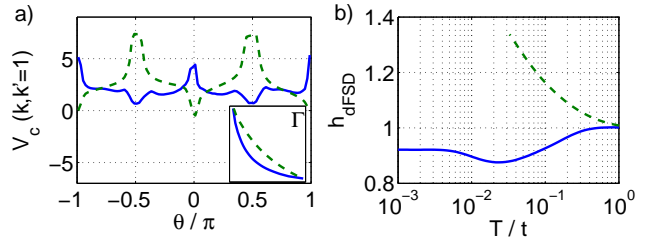}
%\end{center}
\caption{(color online). a): Effective forward scattering in the charge channel $V_c(\vec{k},\vec{k}')$ with $\vec{k}'$ fixed 
near $(-\pi,0)$ ($\theta=-\pi$) obtained with $96$-patch $T$-flow fRG. The inset shows the Fermi surfaces. b): $T$-flow of the 
coupling to static external $d$FSD fields, averaged around the FS. 
In all plots, the solid lines are for $t'=0.44t$, the dashed lines for $t'=0.25t$. 
 }
\label{forwplot}
\end{figure}

The fRG flow is started at high temperatures $T \sim$ bandwidth with pure onsite repulsion $U$. 
In the flow to lower $T$, one-loop corrections generate a detailed  dependence of the interactions $V_T(\vec{k}_1,\vec{k}_2,\vec{k}_3)$ on the wave vectors $\vec{k}_1$, $\vec{k}_2$, and $\vec{k}_3$. 
In the Hubbard model with $U=3t$ near the VH filling, the fRG basically finds two regimes\cite{tflow}. 
One is for $|t'|<|t|/3$, where the leading ordering tendencies are either in the antiferromagnetic channel for strong nesting, or, if the nesting is weaker, in the $d_{x^2-y^2}$-wave pairing channel\cite{tflow}. For for $|t|/3 < |t'|< 1/2$, the leading instability is in the FM 
channel.
With bare onsite interactions the charge channel never becomes leading.
However, e.g., the forward scattering 
$V_c (\vec{k},\vec{k'}) \bar{c}_{\vec{k},s} \bar{c}_{\vec{k}',s'}  {c}_{\vec{k}',s'}
 {c}_{\vec{k},s}$ with zero momentum transfer can develop a pronounced dependence on $\vec{k}$ and $\vec{k'}$. 
This can give rise to subdominant instabilities such as FS deformations\cite{halboth}. 
The suggested $d$-wave FS deformation 
needs a component in  $V_c (\vec{k},\vec{k'})$ of the type $- V_{c,2} (\cos k_x - \cos k_y)\cdot (\cos k_x' - \cos k_y')$ with $V_{c,2}>0$. 
In Fig.~\ref{forwplot} a) we show the fRG result for $V_c (\vec{k},\vec{k'})$ at 
low temperatures above the runaway flow of the leading tendency, obtained with the fRG scheme of Ref.~\onlinecite{tflow}. 
The dashed line is for temperatures above a $d$-wave pairing instability at the VH filling for  $t'=0.25t$ and $T=0.09t$. 
One can clearly observe the attractive $d$-wave component in $V_c (\vec{k},\vec{k'})$ with $\vec{k}'$ fixed at $(-\pi,0)$. 
The solid line is near VH filling and $t'=0.44t$ in the FM regime at $T=0.04t$. Now there is no sign for an attraction in the $d$-wave channel. The couplings to external static and uniform  sources in this channel does not increase for $T \to 0$ (Fig.~\ref{forwplot}~b)). 
These results reflect the influence of the spin fluctuations on the forward 
scattering. Quite generally, the forward scattering  between the parts of the FS  connected  by the dominant spin fluctuation wavevector $\vec{q}$ is renormalized to stronger repulsion. The strong $(\pi,\pi)$-fluctuations between 
the VH points create the $+$-lobe of the $d$-wave in $V_c (\vec{k},\vec{k'})$ in the AF/$d$-wave pairing regime for small $t'$. 
The attraction for $\vec{k} \approx \vec{k}'$ is a precursor of the pairing 
tendencies. In the FM regime, 
the dominant spin fluctuations have $\vec{q} \approx 0$, i.e. make $\vec{k} \approx \vec{k}'$ more repulsive. Hence the $d$-wave charge forward scattering turns out to be slightly repulsive rather than attractive. 
In fact, there is not any  clearly attractive component in 
$V_c (\vec{k},\vec{k'})$ in the FM regime. When we move the FS further away from the VH points by a density change or by a weak Zeeman splitting, the FM instability is cut off. Nevertheless, the main features of the charge forward scattering remain robust.  

We conclude that at least in a single-band scenario with dominant FM fluctuations, it is difficult to identify a microscopic mechanism for a $d$-wave FS deformation. 
Conversely, if we assume that the relevant FS of Sr$_3$Ru$_2$O$_7$  is on the AF/$d$-wave side, the spin response should be strongest near the wavevector $\vec{q} = (\pi,\pi)$ connecting the two VH regions. 
This seems unlikely in view of the strong FM tendencies in all  layered strontium-ruthenates, and in 
inelastic neutron scattering there is no sign for these fluctuations\cite{capogna}.
Furthermore, at least in the Hubbard model, 
the $d$FSD tendencies are always weaker than other instabilities. Hence, we would need a mechanism that suppresses the leading tendencies. 

We now turn to a second scenario for the metamagnetic transition that exploits the vicinity to ferromagnetism. 
This tendency is common all the strontium-ruthenates mentioned in the introduction. 
In contrast with the $d$FSD instability 
it fits readily into the picture drawn by a fRG treatment, as FM is found as the leading instability 
of the weakly-coupled Hubbard model near the VH filling for $1/3< |t'| < 1/2$\cite{tflow}. 
For this situation, the metamagnetic transition was modeled by Binz and Sigrist\cite{binz} using a mean-field theory with a density of states 
$\rho(\epsilon) = \frac{1}{2W} \log [W/|\epsilon - \epsilon_{\mathrm{VH}}|]$ in a band $-W<\epsilon - \epsilon_{\mathrm{VH}}<W$. The electrons interact with an onsite interaction $U>0$.
Their density $n$ and the distance to the VH filling at $n=1$ can be adjusted by $\epsilon_{\mathrm{VH}}$. 
Minimizing the Gibbs potential $G(T,h,n)$ with respect to the magnetization $m$ yields a density region near 
the VH filling at low $T$ where $m$ exhibits a sudden jump at a critical magnetic field $h_m$. 
Coming down from high $T$ at $h_m$, $m$ increases strongly and finally saturates below a temperature scale $T_m$. Impurities cut off the logarithmic growth of the DOS near the VH filling. 
Hence, they are expected to smear out the metamagnetic transition. In this simple model, the magnetization does not 
depend on the direction of the $h$-field, but this effect can be taken into account, e.g., by including anisotropic interactions.
 
\begin{figure} 
%\begin{center} 
\includegraphics[width=.47\textwidth]{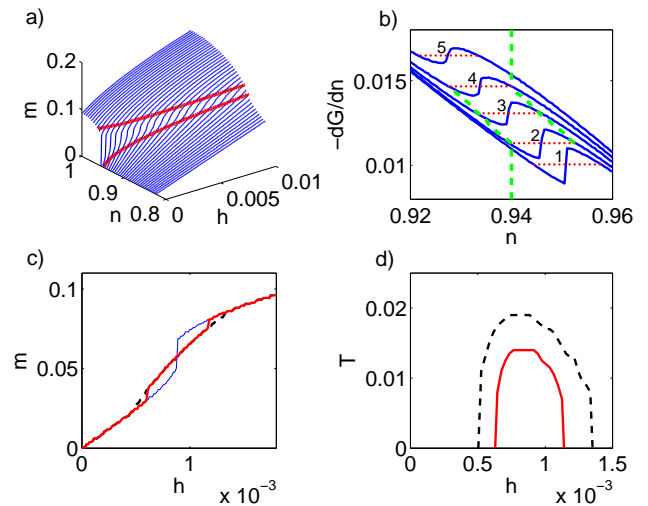}
%\end{center}
\caption{(color online). a): Magnetization $m$ at temperature $T$=0 vs. magnetic field $h$ in units of the half bandwidth $W$ and density $n$, $U=0.4W$.
The two thick lines bound the phase separated region of the Maxwell construction.  
b): $-\mu = - \partial G/\partial n$ for  $h=0.0002 \dots 0.0019W$ with $h$ increasing from curve 1 to 5. The dotted lines indicate the Maxwell construction and span the phase separated region in the upper left plot. The dashed lines show the density components vs. $h$ for average density $n=0.94$. c): $m$ at $T$=0 vs. $h$ for total density $n=0.94$.
The thin line is for the homogeneous case, the dashed line is the Maxwell construction allowing for zero interface energy $G_I=0$. The thicker line with two jumps is for $G_I = 5\cdot 10^{-7}W$.  
d): Boundaries of the inhomogeneity regions for total density $n=0.94$ vs. $h$  and $T$. 
The dotted lined is again for $G_I=0$ and the thick line for $G_I = 5\cdot 10^{-7}W$.
 }
\label{mgfig}
\end{figure}

In Fig.~\ref{mgfig}~a) we show the steplike increase of the magnetization $m$ with increasing magnetic field $h$ for different electron densities $n$, equivalent to the data of Ref.~\onlinecite{binz}. In Fig.~\ref{mgfig}~b) we show the derivative of the Gibbs potential 
$G(T,h,n)$ with respect 
to $n$ versus $n$ for different $h$. For the critical $h$-values $h_m$ where the jump in $m$ occurs, 
the curvature of $G(T,h,n)$ with respect to $n$, corresponding to the inverse compressibility, is negative. 
Hence, at the metamagnetic transition, the system is unstable with respect to phase separation. 
If we plot $G(T,h,n)$ versus $n$  we find a small upward dent in the critical density range. 
In this range we can minimize the energy by the Maxwell construction, i.e., 
by bridging the dent from below with a straight line which lies below  $G(T,h,n)$ between $n_<(h)$ and $n_>(h)$.
 The average density is a mixture between a less dense phase $n_<(h)$ with lower $m$
and a denser phase $n_>(h)$ with higher $m$, the mixing ratio $0< p <1$ is determined by $p n_<(h)+ (1-p) n_>(h) =n$. 
When we increase $h$ from below at a fixed total density $n$, we enter the unstable 
region at a certain lower critical field $h_<(n)$ with $n_<(h_<(n))=n$. 
In Fig.~\ref{mgfig}~b) 
for $n=0.94$ this occurs for $h$ a little higher than $h=0.0005W$ for curve 2. For curve 3 the stable solution is a mixture between $n_<(h)$ and $n_>(h)$. The path of $n_<(h)$ and $n_>(h)$ with increasing $h$ is shown by the dashed lines.
Going up, the weight $p$ of $n_<(h)$ decreases until it vanishes for curve 4. 
In the coexistence region, 
$\mu=\partial G /\partial n$ is constant and the inverse compressibility $\kappa^{-1} \propto \partial^2 G /\partial n^2$ vanishes. 
With increasing $h$, the instability 
region moves to smaller densities, and above an upper critical field $h_>(n)$, 
the homogeneous density $n=0.94$ has 
the lowest Gibbs energy again (curve 5). The wiggle in 
$\partial G/ \partial n$, i.e. the energy difference between the phase separated and the homogeneous solution, becomes smaller with increasing $h$.

Again, this is very similar to a liquid-gas transition. 
There the intensive tuning parameter is the temperature $T$ instead of $h$. 
At the transition, the conjugated extensive quantity, the entropy jumps upwards. 
As a function of the volume $V$, the Gibbs potential has negative curvature around the 
critical volume where the transition occurs, and the isothermal lines in 
the pressure($P=-\partial G/\partial V$)-volume diagram exhibit a similar wiggle as $\mu = \partial G/\partial n$ 
in our case. Hence the inhomogeneous phase at the metamagnetic transition just corresponds to the coexistence regime 
in the liquid-gas transition. 

As can be seen Fig.~\ref{mgfig}~b), in our model the density 
 difference between the two components of the mixture 
is  rather weak of the order $0.01$/site. 
In a more realistic model including long range Coulomb interactions, 
the phase separation will be frustrated on a microscopic length 
scale\cite{lorenzana}. 
Then the two-phase mixture will create a pattern of nano-domains. 
Elongated domains like stripes are a likely candidates. 
Here we speculate that this charge inhomogeneity may be 
responsible for the observed plateau in the residual resistivity at the metamagnetic transition. The multiple Fermi surfaces of Sr$_3$Ru$_2$O$_7$ might help to reduce the Coulomb energy of the charge distribution by mutual screening. 

In the liquid-gas transition the entropy jump is conserved by the Maxwell construction if the transition is crossed at fixed pressure. 
In the metamagnetic case the total density is fixed during the transition. 
The Maxwell construction interpolates between the low magnetization below the transition and the high-$m$ phase above. 
This transforms the step at $h_m$ into a continuous rise between the lower and upper coexistence fields, $h_<$ and $h_>$. 
Across the transition, 
a fraction $0<p<1$ will be already in the dense large-$h$ phase corresponding to $n_>(h)$, while $1-p$ will remain in the low-$h$ less dense phase with $n_<(h)$, and $p$ increases from 0 to 1.
Now, the Coulomb interaction will enforce charge neutrality on a nano scale.  
It is natural to assume that the interface between the two phases and the remaining Coulomb energy due to the inhomogeneity costs some energy $G_I>0$.  
Taking this into account, the straight line for the energy of the phase-separated solution which undercuts the Gibbs potential of the homogeneous system between $n_<(h)$ and  $n_>(h)$ gets shifted upward by $G_I$. 
Then nano-scale phase separation is energetically favorable only for $p$-values between $p_{\mathrm{min}}$ and  $p_{\mathrm{max}}$, otherwise the homogeneous state will be lower in energy. 
When we compute the magnetization with this additional constraint, 
$p_{\mathrm{min}}< p < p_{\mathrm{max}}$, we obtain two steps. One is near $h_<$, where $p$ jumps 
from 0 to a nonzero $p_{\mathrm{min}}$. The other is near $h_>$, where the minority domains are squeezed out, 
and $p$ jumps from $p_{\mathrm{max}}$ to 1. 
Hysteresis is most likely at these two jumps, in between the evolution is a continuous change of the relative size $p$ of the two components and their densities. A precise calculation of $G_I$ is difficult. In order to outline our idea, in Fig.~\ref{mgfig}~c) we assume a value $G_I = 5\cdot 10^{-7}W$ which is about $1/5$ of the maximal energy gain of the inhomogeneous solution. Obviously, if $G_I$ is too large, inhomogeneity does not occur.

In the experiments\cite{grigera04}, the resistivity anomaly becomes smaller when the in-plane component 
of the magnetic field is increased. Just as was suggested for the $d$FSD domains\cite{grigera04}, this could have the effect of 
orienting the density pattern. 
Then the transport along and orthogonal to the in-plane field  should be different.

When the temperature is increased the magnetization jumps get smoothed out. As the spin symmetry is broken by $h\not= 0$, 
there can be no phase transition associated with a symmetry breaking. 
However the instability towards phase separation described above yields 
a thermodynamical distinction between 
the low-$T$ micro-phase separated state and the normal state at higher $T$. 
Above a threshold temperature $T_{\mathrm{neg}}$, 
the negative curvature region in the Gibbs potential disappears. 
Hence the density remains homogeneous through the $h$-range where $m$ rises steeply. 
We speculate that $T_{\mathrm{neg}}$  is an upper bound for the temperature scale below which the $T$-dependent inelastic scattering becomes unobservable in the experiments.  If the interface energy $G_I$ is $T$-independent, the inhomogeneity only 
sets in at $T<T_{\mathrm{neg}}$.
In Fig.~\ref{mgfig}~d) we show the 
inhomogeneity regions in the $h$-$T$ diagram for $G_I=0$ and for a $T$-independent $G_I=5\cdot 10^{-7}W$. Lifetime effects due to impurities act like increasing $T$ 
and reduce the inhomogeneity region as well.

Experimentally, this scenario could be checked by local probes which are able to resolve spatial variations of the charge distribution, e.g. scanning tunneling microscopy. Note that in Ref.~\cite{grigera04} the authors state carefully that phase separation cannot be ruled out near the transition.

In conclusion, we analyzed two possible scenarios for the metamagnetic transition in the layered ruthenates. 
For the $d$-wave FS deformation picture\cite{grigera04,kee}, 
a simple one-band $t$-$t'$ Hubbard model can account for a charge instability 
that splits the saddle points of the dispersion only if the dominant magnetic fluctuations have a large wave 
vector connecting these saddle points. 
This does not seem to be the case in view of the many experimental and 
theoretical indications that strontium-ruthenates are close to ferromagnetism. For dominant FM fluctuations, we do not find an attractive coupling constant 
for the $d$-wave FS deformations. As an alternative we 
extended the Stoner model of itinerant metamagnetism by 
Binz and Sigrist\cite{binz} and analyzed the possibility of 
microscopic phase separation. 
The metamagnetic transition in this model occurs in a region which is unstable with respect to density variations. Due to this general tendency at first order transitions the charge distribution might become inhomogeneous. We discussed the effects of the Coulomb repulsion and interface energies. If they do not cancel the energy gain by the phase separation completely, nano-scale inhomogeneity could account for the observed anomaly in the residual resistivity and the occurrence of two steps 
in the magnetization. 

Finally we note that Binz et al.\cite{metaZH} have considered a scenario with uncharged magnetic domains due to the demagnetization effect. 
We hope that further experiments can distinguish between the various 
proposals. 

I thank B. B\"uchner, A. Katanin, A. Mackenzie, W. Metzner, 
T.M. Rice, M. Sigrist, and H. Yamase for useful communications and comments.

\end{document}